\documentclass[12pt]{article}

\usepackage{array,dsfont} 
\usepackage{epsfig}
\usepackage{amssymb}
\usepackage{graphics,graphpap}

\renewcommand{\thefootnote}{\fnsymbol{footnote}}

\begin{document}

\begin{titlepage}
\begin{flushright}\begin{tabular}{l}
IPPP/06/94\\
DCPT/06/188
\end{tabular}
\end{flushright}
\vskip1.5cm
\begin{center}
   {\Large \bf\boldmath $|V_{td}/V_{ts}|$ from QCD Sum Rules on the Light-Cone}
    \vskip2.5cm {\sc
Patricia Ball\footnote{Patricia.Ball@durham.ac.uk}
}
  \vskip0.5cm
{\em         IPPP, Department of Physics,
University of Durham, Durham DH1 3LE, UK }\\
\vskip2.5cm 


\vskip2.5cm

{\large\bf Abstract\\[10pt]} \parbox[t]{\textwidth}{
The dominant theoretical uncertainty in extracting $|V_{td}/V_{ts}|$
from the ratio of branching ratios $R_{\rho/\omega}\equiv\bar{\cal B}(B\to
(\rho,\omega)\gamma)/\bar{\cal B}(B\to
K^* \gamma)$ is given by the ratio of form factors $\xi_{\rho}\equiv T_1^{B\to
  K^*}(0)/T_1^{B\to \rho}(0)$. We find $\xi_{\rho}= 1.17\pm 0.09$ from
QCD sum rules on the light-cone. Using QCD factorisation for the 
branching ratios, including the most dominant power-suppressed effects
beyond QCD factorisation, and the current experimental results for
$R_{\rho/\omega}$, this translates into
$|V_{td}/V_{ts}|_{\rm BaBar} = 0.199^{+0.023}_{-0.025}({\rm exp})\pm 
0.014 ({\rm th})$, which corresponds to $\gamma_{\rm BaBar} = 
(61.0^{+13.5}_{-16.0}({\rm   th}){}^{+8.9}_{-9.3}({\rm th}))$, and 
$|V_{td}/V_{ts}|_{\rm Belle} = 0.207^{+0.028}_{-0.033}({\rm exp})
^{+0.014}_{-0.015}({\rm th})$, $\gamma_{\rm Belle} = 
(65.7^{+17.3}_{-20.7}({\rm exp})^{+8.9}_{-9.2}({\rm th}))^\circ$.

}
\vskip0.5cm

$$\epsfxsize=0.4\textwidth\epsffile{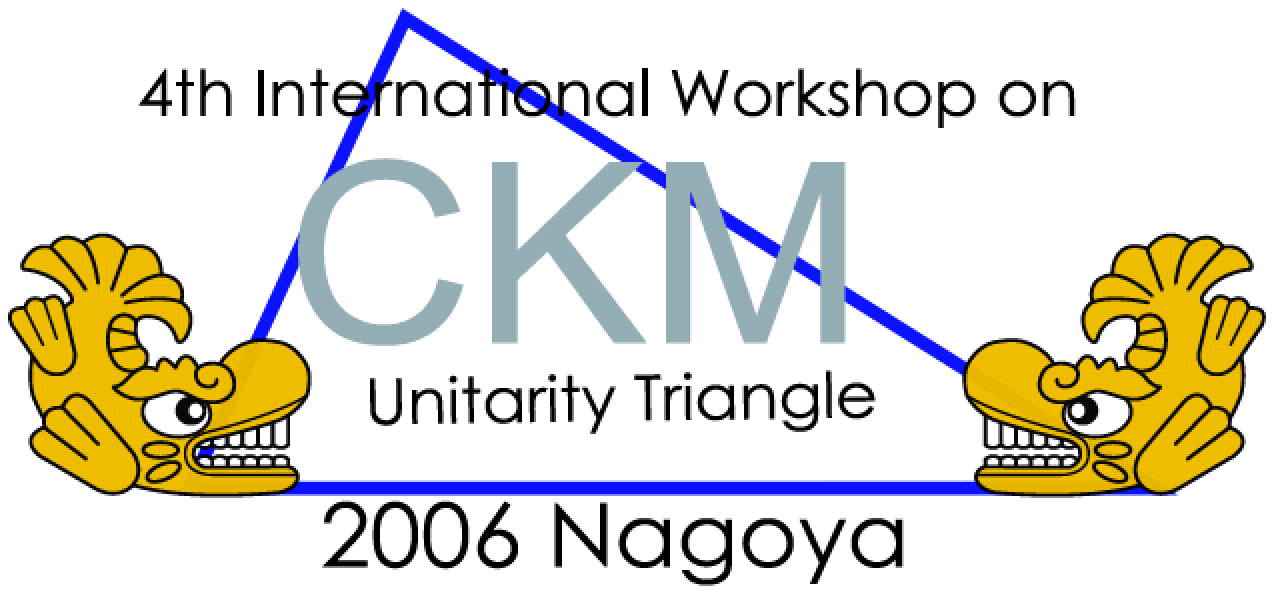}$$

\centerline{\em Talk given at CKM06, Nagoya (Japan), Dec 2006}
\end{center}

\vfill

\end{titlepage}

\setcounter{footnote}{0}
\renewcommand{\thefootnote}{\arabic{footnote}}

\newpage

Both BaBar and Belle have seen the $b\to d$
penguin-dominated decays $B\to(\rho\,,\omega)\gamma$ \cite{Babar,Belle}.
Assuming the Standard
Model (SM) to be valid, these processes offer the possibility to extract the
CKM matrix element $|V_{td}|$, in complementarity to the
determination from $B_d$ mixing and the SM unitarity triangle
based on $|V_{ub}/V_{cb}|$ and the angle $\gamma$. 
In order to extract $|V_{td}|$ from
the measured rate, one needs to know both short-distance weak and strong
interaction effects and long-distance QCD effects. Whereas the former
can, at least in principle, 
be calculated to any desired precision in the framework of
effective field theories, and actually  are currently known to
(almost) NNLO in
QCD \cite{SD}, the assessment of long-distance QCD effects has been 
notoriously difficult. A solution to this problem is provided
by QCD factorisation (QCDF) 
\cite{BVga,BoBu}, a consistent framework allowing one to
write the relevant hadronic matrix elements as
\begin{equation}\label{1}
\langle V\gamma|Q_i| B\rangle =
\left[ T_1^{B\to V}(0)\, T^I_{i} +
\int^1_0 d\xi\, du\, T^{II}_i(\xi,u)\, \phi_B(\xi)\, \phi_{V;\perp}(v)\right]
\cdot\epsilon\,.
\end{equation}
Here $\epsilon$ is the photon polarisation 4-vector, $Q_i$ is one of
the operators in the  effective Hamiltonian,
$T_1^{B\to V}$ is a $B\to V$ transition form factor,
and $\phi_B$, $\phi_{V;\perp}$ 
are leading-twist light-cone distribution amplitudes (DAs)
of the $B$ meson and the vector meson $V$, respectively.
These quantities are universal non-perturbative objects and
describe the long-distance dynamics of the matrix elements, which
is factorised from the perturbative short-distance interactions
included in the hard-scattering kernels $T^I_{i}$ and $T^{II}_i$.
The above QCDF formula is valid in the heavy-quark limit
$m_b\to\infty$ and is subject to 
corrections of order $\Lambda_{\rm QCD}/m_b$. Although it is possible
to determine $|V_{td}|$ from the branching ratio of
$B\to(\rho,\omega)\gamma$ itself, the associated theoretical
uncertainties get greatly reduced when one considers the ratio of
branching ratios for $B\to K^*\gamma$ and $B\to (\rho,\omega)\gamma$
instead. One then can extract $|V_{td}/V_{ts}|$ from
\begin{equation}\label{Brat}
\frac{\bar{\cal B}(B\to(\rho,\omega)\gamma)}{\bar{\cal B}(B\to K^*\gamma)} =
\left|\frac{V_{td}}{V_{ts}}\right|^2
\left(\frac{1-m_{\rho,\omega}^2/m_B^2}{1-m_{K^*}^2/m_B^2}\right)^3
\left( \frac{T_1^{\rho,\omega}(0)}{T_1^{K^*}(0)}\right)^2 \left [ 1 +
  \Delta R\right];
\end{equation}
$\Delta R$ contains all non-factorisable effects
induced by $T_i^{I,II}$ in (\ref{1}). 
The theoretical uncertainty of
this determination is governed by both the ratio of form factors
$T_1^{K^*}(0)/T_1^{\rho,\omega}(0)$ and the value of $\Delta R$, which
parametrises not only SU(3)-breaking effects, but also power-suppressed
corrections to QCDF. In this talk, I report
the results of a recent calculation of both the form factor ratios and
the $\Delta R$ term, including effects beyond QCDF, see 
Refs.~\cite{PB3,PB6,PB8}.

We have calculated $T_1$ previously, in Refs.~\cite{BZ04}, using the
method of QCD sum rules on the light-cone.
Here, we focus on the ratios
\begin{equation}\label{xi}
\xi_\rho \equiv \frac{T_1^{B\to K^*}(0)}{T_1^{B\to \rho}(0)}\,, \quad
\xi_\omega \equiv \frac{T_1^{B\to K^*}(0)}{T_1^{B\to \omega}(0)}\,,
\end{equation}
which govern the extraction of $|V_{ts}/V_{td}|$ from $B\to V\gamma$
decays. Compared with our previous results of Ref.~\cite{BZ04}, 
we implement the following improvements:
\begin{itemize}
\item updated values of SU(3)-breaking in twist-2 parameters \cite{elena,PB1};
\item complete account of SU(3)-breaking in twist-3 and -4 DAs \cite{prep};
\item NLO evolution for twist-2 parameters.
\end{itemize}
The sum rules can of course be used to determine each form factor
separately, but it turns out that the ratio is more accurate, because 
$\xi_{\rho,\omega}$ is independent of the $B$-meson decay constant $f_B$
and also, to very good accuracy, of $m_b$ and the sum rule parameters 
$M^2$ and $s_0$, i.e.\ a good part of the systematic uncertainties of
the method cancel. However, $\xi_{\rho,\omega}$ is
very sensitive to SU(3)-breaking effects in the DAs, and 
it is precisely these effects we shall focus on in this paper. A
similar analysis for the ratio of the $D\to K$ and $D\to\pi$ form
factors was carried
out in Ref.~\cite{0608}.
\begin{figure}[p]
$$
\epsfxsize=0.45\textwidth\epsffile{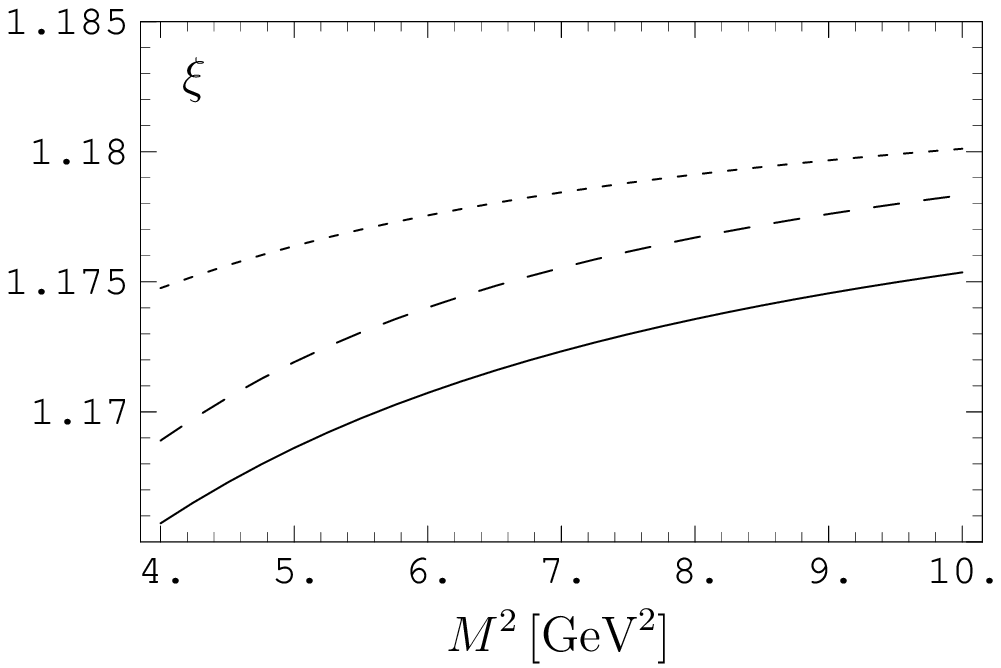}\qquad
\epsfxsize=0.45\textwidth\epsffile{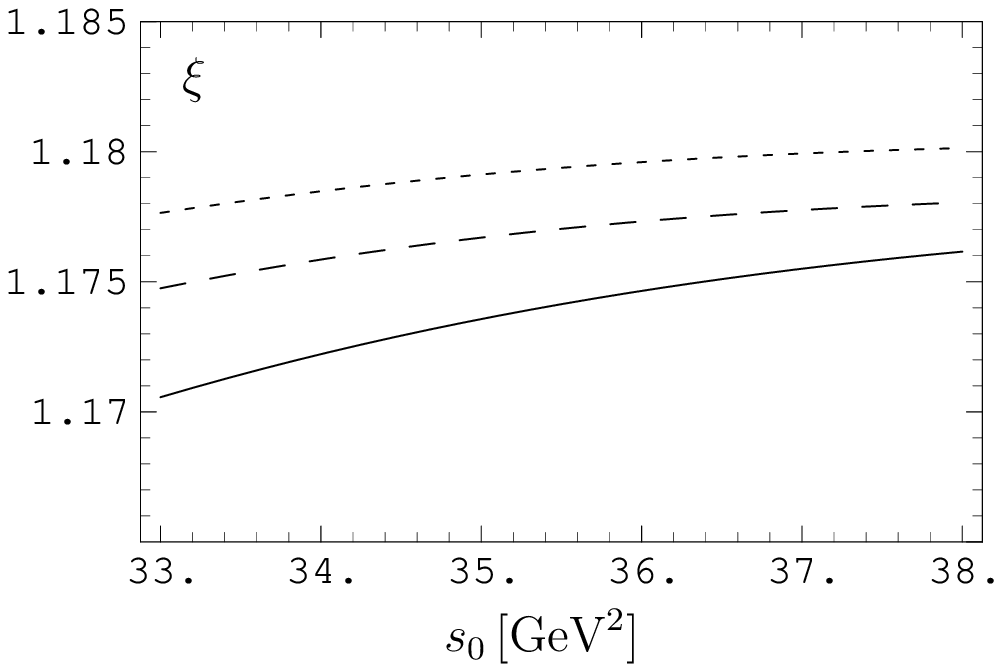}$$
\vspace*{-30pt}
\caption[]{Left panel: $\xi_\rho$ as a function of the Borel parameter $M^2$
  for $s_0 = 35\,{\rm GeV}^2$ and central values of the input
  parameters. Right panel: $\xi_\rho$ as a function of the continuum
  threshold $s_0$ 
  for $M^2 = 8\,{\rm GeV}^2$ and central values of the input
  parameters. Solid lines: DAs in conformal expansion; long dashes: BT
  model \cite{angi} for twist-2 DAs; short dashes: BT model for
  twist-2 DAs and 
renormalon model for twist-4 DAs \cite{renormalon}.}\label{fig:1}
$$
\epsfxsize=0.45\textwidth\epsffile{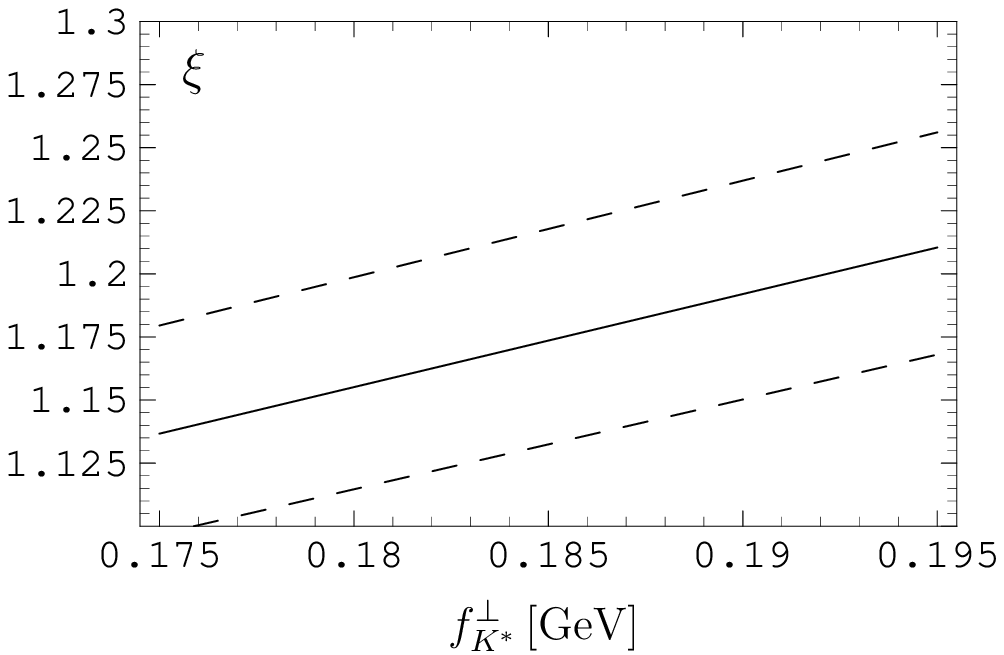}
$$
\vspace*{-30pt}
\caption[]{$\xi_\rho$ as a function of $f_{K^*}^\perp(1\,{\rm GeV})$. Solid
  line: $f_{\rho}^\perp(1\,{\rm GeV})=0.165\,{\rm GeV}$, dashed lines:
  $f_{\rho}^\perp$ shifted by $\pm 0.009\,$GeV.}\label{fig:2}
$$
\epsfxsize=0.45\textwidth\epsffile{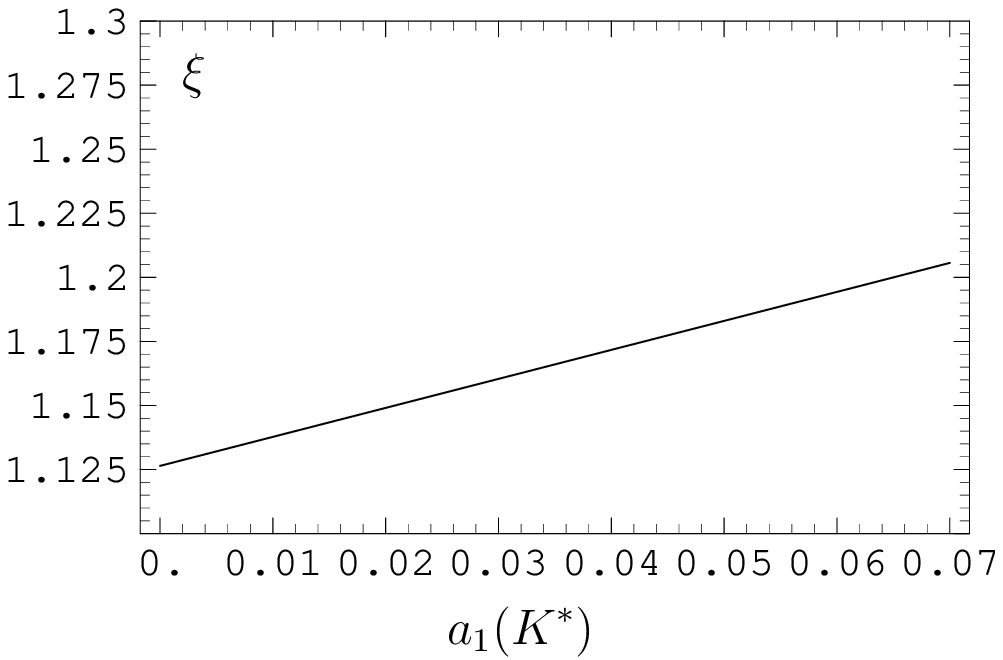}\qquad
\epsfxsize=0.45\textwidth\epsffile{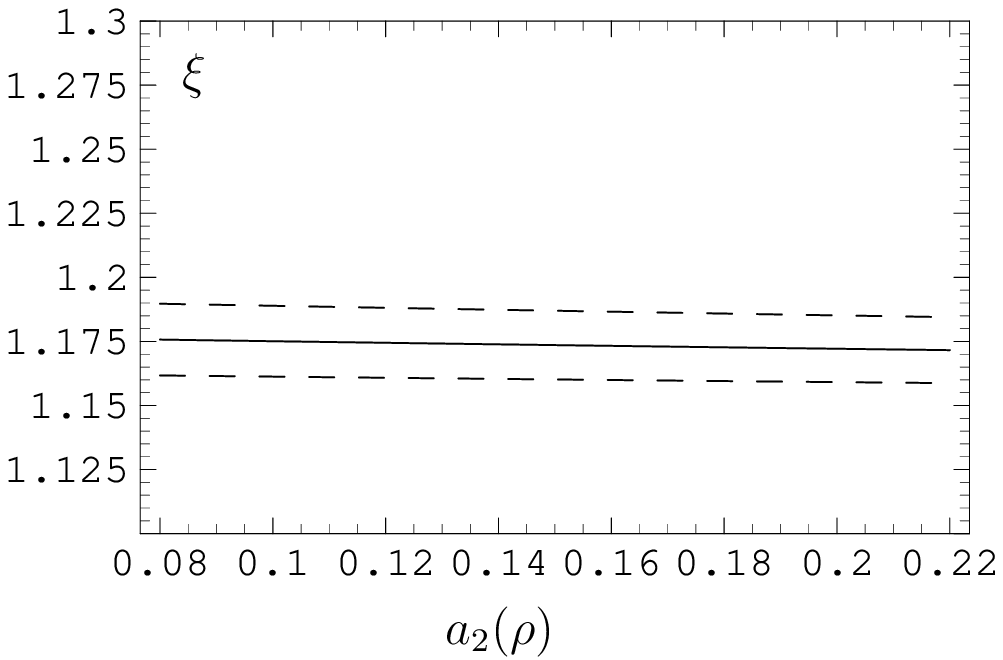}$$
\vspace*{-30pt}
\caption[]{Left panel: $\xi_\rho$ as a function of $a_1(K^*)$ at
  1~GeV. Right panel: $\xi_\rho$ as a function of $a_2(\rho)$ at 1~GeV. 
  Solid line: $a_2(K^*) = a_2(\rho)-0.04$; dashed lines: $a_2(K^*)$
  shifted by $\pm 0.02$. Longitudinal and transverse parameters
  $a_i^\parallel$ and $a_i^\perp$ are
  set equal.}\label{fig:3}
\end{figure}

Let us now discuss the results for $\xi_\rho$ and
their uncertainty. The light-cone 
sum rule for $\xi_\rho$, Ref.~\cite{PB3}, depends
on sum-rule specific parameters, the Borel parameter $M^2$ and the
continuum threshold $s_0$, and on hadronic input parameters, i.e.\
decay constants $f_{\rho,K^*}^{\parallel,\perp}$, Gegenbauer moments of 
twist-2 light-cone DAs of the $\rho$ and $K^*$,
$a_{1,2}^{\parallel,\perp}(\rho,K^*)$, 
and parameters describing twist-3 and 4 DAs. 
In Fig.~\ref{fig:1} we plot the dependence of $\xi_\rho$
on the Borel parameter and the continuum threshold. The curves are
very flat, which indicates that the systematic
uncertainties of the light-cone sum rule approach cancel to a
large extent in the ratio of SU(3)-related form factors. In
Fig.~\ref{fig:2} we plot $\xi_\rho$ as a function of
$f_{K^*}^\perp$, for various values of
$f_{\rho}^\perp$. The uncertainty in both parameters
causes an uncertainty in $\xi_\rho$ of $\pm 0.08$. In Fig.~\ref{fig:3},
left panel, we show the dependence of $\xi_\rho$ on $a_1(K^*)$, which
induces a change in $\xi_\rho$ by $\pm 0.03$. 
The right panel shows the dependence on $a_2$
which is rather mild and causes $\xi_\rho$ to change by $\pm 0.02$. The
variation of the remaining parameters within their respective 
limits causes another $\pm 0.02$ shift in $\xi_\rho$, so that we
arrive at the following result \cite{PB3}:
\begin{eqnarray}
\xi_\rho = \frac{T_1^{B\to K^*}(0)}{T_1^{B\to \rho}(0)} &=& 1.17 \pm
0.08(f^\perp_{\rho,K^*}) \pm 0.03(a_1)
\pm 0.02(a_2) \pm 0.02(\mbox{twist-3 and -4})\nonumber\\
&&{} \pm 0.01 (\mbox{sum-rule
  parameters, $m_b$ and twist-2 and -4 models})\nonumber\\
&= &1.17\pm 0.09\,.\label{resxi}
\end{eqnarray}
The total uncertainty of $\pm 0.09$ is obtained by adding the 
individual terms in quadrature. $\xi_\rho$ was also obtained from a
quenched lattice calculation \cite{mescia}: $\xi_{\rho,{\rm latt}} =
1.2\pm 0.1$, which agrees with our result. 
An analogous calculation of
$\xi_\omega$ yields \cite{PB8}:
\begin{equation}\label{xiomega}
\xi_\omega \equiv \frac{T_1^{B\to K^*}(0)}{T_1^{B\to \omega}(0)} = 1.30\pm
0.10\,.
\end{equation}

Let us now turn to the calculation of the ratio of branching ratios
and the determination of $|V_{td}/V_{ts}|$. BaBar and Belle
have measured the quantity 
$$
R_{\rho/\omega} \equiv 
\frac{\overline{\cal B}(B\to (\rho,\omega)\gamma)}{\overline{\cal B}(B\to
  K^*\gamma)}\,,
$$
where $\overline{\cal B}(B\to (\rho,\omega)\gamma)$ is defined as the
CP-average $\frac{1}{2}[{\cal B}(B\to (\rho,\omega)\gamma)+{\cal
  B}(\bar B\to (\bar\rho,\omega)\gamma)]$ of 
$${\cal B}(B\to (\rho,\omega)\gamma) = \frac{1}{2}\left\{ {\cal B}(B^+\to
\rho^+\gamma) + \frac{\tau_{B^+}}{\tau_{B^0}}\left[ {\cal B}(B^0\to
\rho^0\gamma) + {\cal B}(B^0\to\omega\gamma)\right]\right\},
$$
and $\overline{\cal B}(B\to K^*\gamma)$ is the isospin- and
  CP-averaged branching ratio of the $B\to K^*\gamma$ channels. 
The experimental results are \cite{Babar,Belle}
\begin{equation}\label{RexBelle}
R_{\rm exp}^{\rm BaBar}= 0.030\pm 0.006\,,\quad
R_{\rm exp}^{\rm Belle}= 0.032\pm 0.008\,.
\end{equation}

As for the theoretical prediction of $R_{\rho,\omega}$, it turns out
that the exclusive $B\to V\gamma$ process is actually described by two
physical amplitudes, one for each polarisation of the photon:
\begin{equation}\label{4}
\bar{\cal A}_{L(R)} = {\cal A}(\bar B\to V
\gamma_{L(R)})\,,  \qquad
{\cal A}_{L(R)} = {\cal A}(B\to \bar V \gamma_{L(R)})\,,
\end{equation}
where $\bar B$ denotes a $(b\bar q)$ and $V$ a $(D\bar q)$ bound
state.
In the notation introduced in Ref.~\cite{BoBu} in the context
of QCDF, the
decay amplitudes can be written as
\begin{eqnarray}
\bar{\cal A}_{L(R)} &=& \frac{G_F}{\sqrt{2}}\,\left( \lambda_u^D
a_{7}^u( V\gamma_{L(R)}) +
\lambda_c^D a_{7}^c( V\gamma_{L(R)})\right) \langle V
\gamma_{L(R)} | Q_7^{L(R)} | \bar B\rangle
\nonumber\\
&\equiv& \frac{G_F}{\sqrt{2}}\,\left( \lambda_u^D
a_{7L(R)}^u( V) + \lambda_c^D a_{7L(R)}^c( V)\right) \langle V
\gamma_{L(R)} | Q_7^{L(R)} | \bar B\rangle\,,\label{ME}
\end{eqnarray}
and analoguously for ${\cal A}_{L(R)}$. The $\lambda_q^D$, $D=s,d$,
are products of CKM matrix elements.
The
  $a_7^{c,u}$ calculated in Refs.~\cite{BoBu}, coincide,
to leading order in  $1/m_b$, with our
  $a_{7L}^U$, whereas $a_{7R}^U$ are set zero in
  \cite{BoBu}. Our
expression (\ref{ME}) is purely formal and does not imply that the
$a_{7R(L)}^{U}$ factorise at order $1/m_b$. As a matter of fact,
they don't.
The operators $Q_7^{L(R)}$ are given by
$$
Q_7^{L(R)} = \frac{e}{8\pi^2}\, m_b \bar D \sigma_{\mu\nu}
             \left(1 \pm \gamma_5\right)b F^{\mu\nu}
$$
and generate left- (right-) handed photons in the decay
$b\to D\gamma$.
We split the factorisation coefficients into three separate
contributions:
\begin{eqnarray}
a_{7L}^U( V) &=& a_{7L}^{U,{\rm QCDF}}( V) + a_{7L}^{U,{\rm ann}}(
 V) + a_{7L}^{U,{\rm soft}}( V)+\dots\,,\nonumber\\
a_{7R}^U( V) &=& a_{7R}^{U,{\rm QCDF}}( V) + a_{7R}^{U,{\rm ann}}(V)+ 
a_{7R}^{U,{\rm soft}}( V)+\dots\,,\label{asplit}
\end{eqnarray} 
where $a_{7L}^{U,{\rm QCDF}}$ is the leading term in the $1/m_b$
expansion; all other terms are suppressed by at least one power of
$m_b$. We only include those
power-suppressed terms that are either numerically large or relevant
for certain observables. The dots
denote terms of higher order in $\alpha_s$ and
further $1/m_b$ corrections to QCDF, most of which are uncalculable.
The superscript ``ann'' denotes the contributions from weak
annihilation diagrams which are particularly relevant for $B\to
(\rho,\omega)\gamma$. At order $1/m_b$, they can be calculated in QCDF
themselves, but there are additional large corrections of $O(1/m_b^2)$
induced by long-distance photon emission from soft quarks, see
Refs.~\cite{kivel,emi}. We have included these corrections in
Ref.~\cite{PB8}, as well as the ``soft'' contributions induced by
soft-gluon emission from quark loops. 

In terms of these coefficients, and the appropriate CKM parameters,
the non-factorisable correction in Eq.~(\ref{Brat}) can be expressed as
\begin{eqnarray}
1+\Delta R & = & \left|
  \frac{a_{7L}^c(\rho)}{a_{7L}^c(K^*)}\right|^2 \left( 1 +
  {\rm Re}\,(\delta a_\pm + \delta a_0) \left[\frac{R_b^2 - R_b
  \cos\gamma}{1-2 R_b \cos\gamma + R_b^2}\right]\right.\nonumber\\
& & \left. + \frac{1}{2}\left( |\delta a_\pm|^2 + |\delta a_0|^2\right)
  \left\{ \frac{R_b^2}{1-2 R_b \cos\gamma + R_b^2}\right\} \right)
\label{delR}
\end{eqnarray}
with $\delta a_{0,\pm}=
a_{7L}^u(\rho^{0,\pm})/a_{7L}^c(\rho^{0,\pm})-1$. Here $\gamma$ is
one of the angles of the UT ($\gamma = {\rm arg}\, V_{ub}^*$ in the
standard Wolfenstein parametrisation of the CKM matrix) 
and $R_b$ one of its sides:
$$
R_b =
\left(1-\frac{\lambda^2}{2}\right)\frac{1}{\lambda}
\left|\frac{V_{ub}}{V_{cb}}\right|.
$$
\begin{figure}[tb]
$$\epsfxsize=0.45\textwidth\epsffile{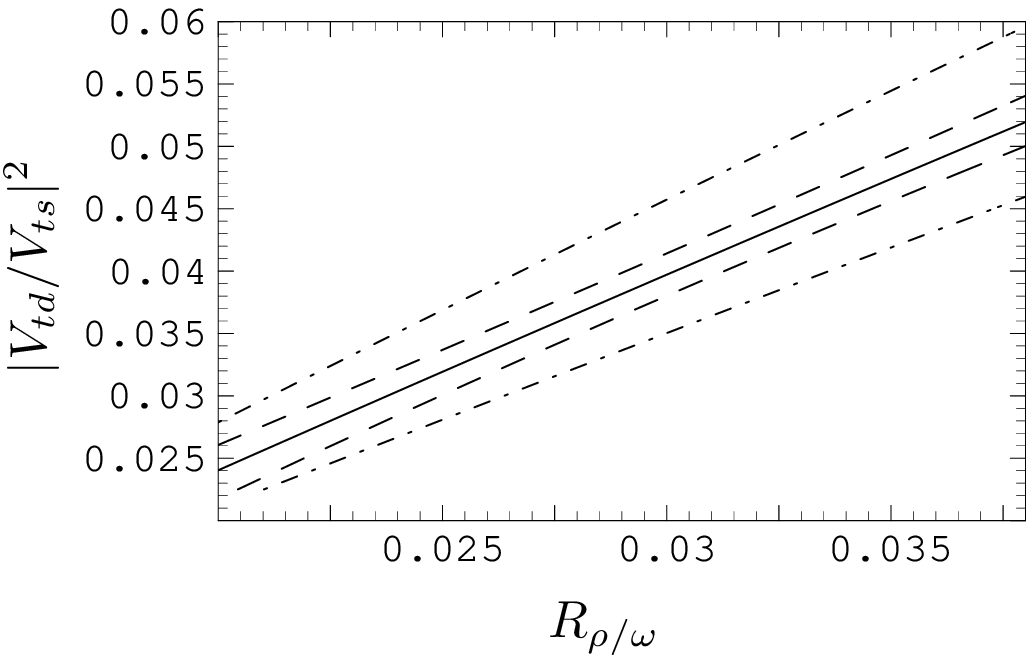}\qquad
\epsfxsize=0.45\textwidth\epsffile{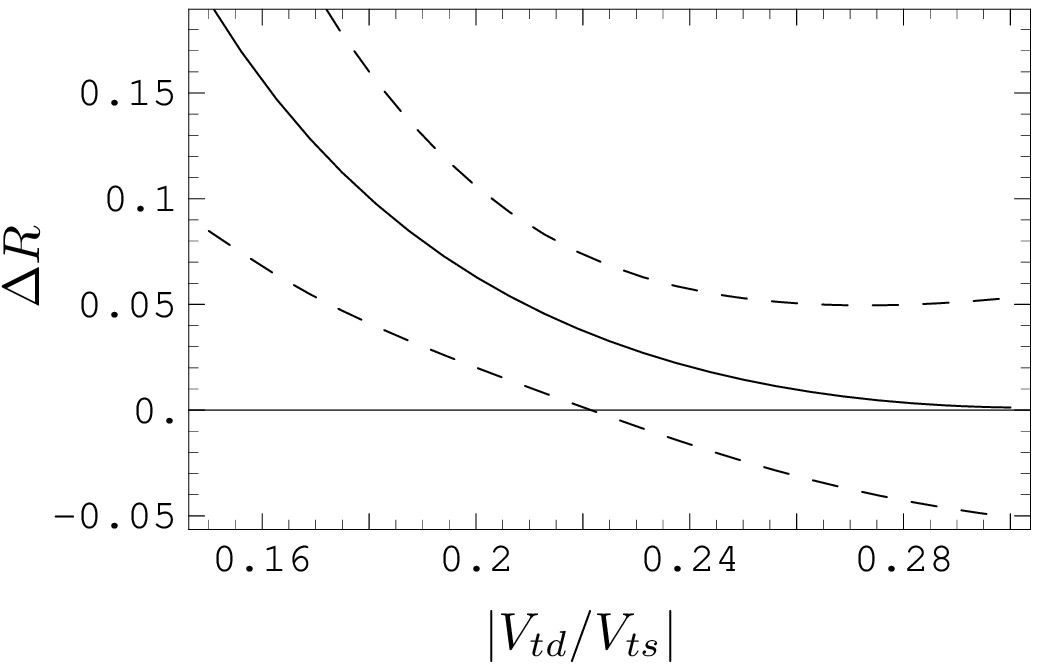}
$$
\vspace*{-25pt}
\caption[]{\small Left panel: $|V_{td}/V_{ts}|^2$ as function of
  $R_{\rho/\omega}$.  
Solid line: central values. Dash-dotted lines: theoretical
  uncertainty induced by $\xi_\rho = 1.17\pm 0.09$.
Dashed lines: other
  theoretical uncertainties. Right panel: $\Delta R$ from 
 Eq.~(\ref{delR}) as function of
  $|V_{td}/V_{ts}|$. Solid
  line: central values. Dashed
  lines: theoretical uncertainty.}\label{fig4}
$$\epsfxsize=0.45\textwidth\epsffile{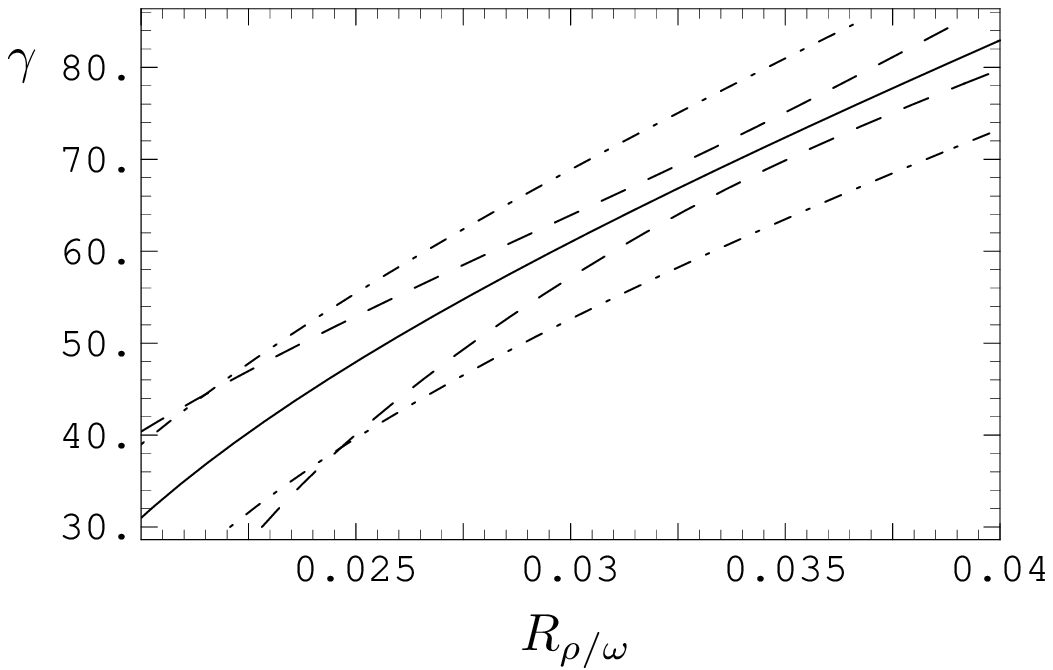}$$
\vspace*{-25pt}
\caption[]{\small The UTangle $\gamma$ as function of 
 $R_{\rho/\omega}$. Solid
  lines: central values of input parameters. Dash-dotted lines: theoretical
  uncertainty induced by $\xi_\rho = 1.17\pm 0.09$. Dashed lines: other
  theoretical uncertainties.}\label{fig5}
\end{figure}
In Figs.~\ref{fig4} and \ref{fig5} we plot the
values of $|V_{td}/V_{ts}|^2$ and $\gamma$, respectively, 
determined from (\ref{Brat}) as a  function of $R_{\rho/\omega}$.
Although the curve in Fig.~\ref{fig4}(a) looks like a straight line,
as naively expected from (\ref{Brat}),
this is not exactly the case, because of the dependence of $\Delta R$
on $|V_{td}/V_{ts}|$. In  Fig.~\ref{fig4}(b) we plot $\Delta R$ as a
function of $|V_{td}/V_{ts}|$. 
The dependence of $\Delta R$ on $|V_{td}/V_{ts}|$ is
rather strong.
 
It is now basically a matter of choice whether to use
$R_{\rho/\omega}$ to determine $|V_{td}/V_{ts}|$ or $\gamma$. Once one of
these parameters is known, the other one follows from
\begin{equation}\label{61}
\left|\frac{V_{td}}{V_{ts}}\right| = \lambda \sqrt{1-2 R_b \cos\gamma
  + R_b^2} \left[ 1 + \frac{1}{2}\,( 1 - 2 R_b \cos\gamma) \lambda^2 +
  O(\lambda^4)\right]\,.
\end{equation} 
In Fig.~\ref{fig5} we plot $\gamma$ as a function of
$R_{\rho/\omega}$, together with the theoretical uncertainties.
In order to facilitate
the extraction of $|V_{td}/V_{ts}|$ (or $\gamma$) from measurements of
$R_{\rho/\omega}$, Tab.~\ref{tabx}
contains explicit values for the theoretical uncertainties for
representative values of $R_{\rho/\omega}$. The uncertainty
induced by $\xi_{\rho}$ is dominant. As discussed in
Ref.~\cite{PB3}, a reduction of this uncertainty
 would require a reduction of the
uncertainty of the transverse decay constants $f_V^\perp$ of $\rho$
and $K^*$. With the most recent results from BaBar, $R_{\rho/\omega} =
0.030\pm 0.006$ \cite{Babar}, and from Belle, $R_{\rho/\omega} =
0.032\pm 0.008$ \cite{Belle}, we then find
\begin{equation}
\renewcommand{\arraystretch}{1.3}
\begin{array}[b]{l@{\quad}l@{\quad\leftrightarrow\quad}l}
\mbox{BaBar:} & \displaystyle
\left|\frac{V_{td}}{V_{ts}}\right| = 0.199^{+0.022}_{-0.025}({\rm exp})\pm
0.014({\rm th}) &\displaystyle
\gamma = (61.0^{+13.5}_{-16.0}({\rm exp})^{+8.9}_{-9.3}
({\rm th}))^\circ\,,\\[10pt]
\mbox{Belle:} & \displaystyle
\left|\frac{V_{td}}{V_{ts}}\right| = 0.207^{+0.028}_{-0.033}({\rm exp})
^{+0.014}_{-0.015}({\rm th}) &\displaystyle
\gamma = (65.7^{+17.3}_{-20.7}({\rm exp})^{+8.9}_{-9.2}({\rm th}))^\circ\,.
\end{array}
\label{63}
\end{equation} 
\begin{table}[tb]
\renewcommand{\arraystretch}{1.3}
\addtolength{\arraycolsep}{3pt}
$$
\begin{array}{c||c|c|c||c|c|c}
R_{\rho/\omega} & |V_{td}/V_{ts}| & \Delta_{\xi_\rho} & \Delta_{\rm
  other\,th}
& \gamma & \Delta_{\xi_\rho} & \Delta_{\rm
  other\,th}\\\hline
0.026 & 0.183 & \pm 0.012 & \pm 0.007 & 50.8 & {}^{+7.5}_{-8.2} &
  \pm 5.8 \\
0.028 & 0.191 & {}^{+0.012}_{-0.013} & \pm 0.006 & 56.0 &
  {}^{+7.7}_{-8.3} & \pm 4.7\\
0.030 & 0.199 & \pm 0.013 & \pm 0.006 & 61.0 & {}^{+7.9}_{-8.4} &
  \pm 4.0\\
0.032 & 0.207 & {}^{+0.013}_{-0.014} & \pm 0.006 & 65.7 &
  {}^{+8.1}_{-8.5} & \pm 3.6\\
0.034 & 0.214 & \pm 0.014 & \pm 0.006 & 70.2 & {}^{+8.4}_{-8.8} & 
\pm 3.5\\
0.036 & 0.221 & {}^{+0.014}_{-0.015} & \pm 0.006 & 74.5 &
  {}^{+8.8}_{-9.0} & \pm 3.7
\end{array}
$$
\vspace*{-0pt}
\caption[]{\small Central values and uncertainties of
  $|V_{td}/V_{ts}|$ and $\gamma$ extracted from representative values of 
$R_{\rho/\omega}$. 
$\Delta_{\xi_\rho}$ is the uncertainty induced by
$\xi_\rho$ and $\Delta_{\rm other\,th}$ that by other input
  parameters, including $\xi_\omega$ and $|V_{ub}|$.}\label{tabx}
\end{table}
These numbers  compare well with the Belle result  \cite{Bellegamma} 
from tree-level processes, $\gamma=(53\pm 20)^\circ$ and results from  
global fits. We also would like to point out that the above
determination of $\gamma$ is actually a determination of
$\cos\gamma$, via Eq.~(\ref{61}), and implies, in principle, a twofold
degeneracy $\gamma\leftrightarrow 2\pi-\gamma$. This is in contrast to the
determination from $B\to D^{(*)} K^{(*)}$ in \cite{Bellegamma}, which
carries a twofold degeneracy
$\gamma \leftrightarrow \pi+\gamma$. Obviously these two
determinations taken together remove the degeneracy and 
select $\gamma\approx 55^\circ<180^\circ$. If 
$\gamma\approx 55^\circ+180^\circ$ instead, one would have 
$|V_{td}/V_{ts}|\approx 0.29$ from
(\ref{61}), which is definitely ruled out by data. Hence, the result
(\ref{63}) confirms the SM interpretation of $\gamma$ from 
the tree-level CP asymmetries in $B\to D^{(*)} K^{(*)}$.

\subsection*{Acknowledgements}
This work was supported in part by the EU networks
contract Nos.\ MRTN-CT-2006-035482, {\sc Flavianet}, and
MRTN-CT-2006-035505, {\sc Heptools}.

\end{document}